\documentclass[aps,pre,showpacs,preprint]{revtex4}

\usepackage{amsmath}
\usepackage{amssymb}
\usepackage[utf8]{inputenc}
\usepackage{graphicx}
\usepackage{dcolumn}
\usepackage[mathscr]{euscript}
\usepackage{bm}
 
\DeclareMathOperator\erfi{erfi}

\begin{document}

\title{Kinetic theory of a confined quasi-one-dimensional gas of hard disks}

\author{M. Mayo}
\affiliation{F\'{\i}sica Te\'{o}rica, Universidad de Sevilla,
Apartado de Correos 1065, E-41080, Sevilla, Spain}

\author{J. Javier Brey}
\email{brey@us.es}
\affiliation{F\'{\i}sica Te\'{o}rica, Universidad de Sevilla,
Apartado de Correos 1065, E-41080, Sevilla, Spain}
\affiliation{Institute Carlos I for Theoretical and Computational Physics. Facultad de Ciencias. Universidad de Granada, E-18071, Granada, Spain}

\author{M. I. Garc\'{\i}a de Soria}
\email{gsoria@us.es}
\affiliation{F\'{\i}sica Te\'{o}rica, Universidad de Sevilla,
Apartado de Correos 1065, E-41080, Sevilla, Spain}
\affiliation{Institute Carlos I for Theoretical and Computational Physics. Facultad de Ciencias. Universidad de Granada, E-18071, Granada, Spain}

\author{P. Maynar}
\email{maynar@us.es}
\affiliation{F\'{\i}sica Te\'{o}rica, Universidad de Sevilla,
Apartado de Correos 1065, E-41080, Sevilla, Spain}
\affiliation{Institute Carlos I for Theoretical and Computational Physics. Facultad de Ciencias. Universidad de Granada, E-18071, Granada, Spain}

\date{\today }

\begin{abstract}
A dilute gas of hard disks confined between two straight parallel lines is considered. The distance between the two boundaries is in between one and two particle diameters, so that the system is quasi-one-dimensional. A Boltzmann-like kinetic equation, that takes into account the limitation in the possible scattering angles,  is derived. It is shown that the equation verifies an $H$ theorem implying a monotonic approach to equilibrium. The implications of this result are discussed, and the equilibrium properties are derived. Closed equations describing how the kinetic energy is transferred between the degrees of freedom parallel and perpendicular to the boundaries are derived for states that are homogeneous along the direction of the boundaries.  The theoretical predictions agree with results obtained by means of molecular dynamics simulations.

\end{abstract}

\maketitle

\section{Introduction}
\label{s1}
In recent years, the study of transport phenomena in gases or liquids confined in spaces whose characteristic length is comparable to the molecular size, has attracted a lot of attention. This has been prompted and stimulated by the relevant new technological applications of nanofluidics \cite{SHyR08,ByC10}. The experimental  advances ask for a better understanding, at a conceptual level, of the effects that strong confinement has on the non-equilibrium behavior of fluids. Because under these conditions the particles do not explore a bulk-like environment, and because of the asymmetry generated by  the confining boundaries, strongly confined  systems exhibit inhomogeneity and anisotropy, that have both a great impact on their macroscopic  properties. 

Most of the studies carried out up to now on transport in  confined fluids consider, more or less explicitly, that hydrodynamics holds in all the directions, i.e. it is supposed that the length characterizing the confinement is larger that the characteristic hydrodynamic length. Then, for instance, to study diffusion in a fluid that is confined between two parallel plates, the starting point is the three-dimensional diffusion equation, with the appropriate boundary conditions confining the system \cite{Zw92,RyR01,MyM11,MMyP15}. In more general transport problems, some {\em ad hoc} extrapolation of the Navier-Stokes equations to confining geometries are employed. In these works, the theoretical problem is how to project the three-dimensional dynamics on a one-dimensional or two-dimensional space, depending on the specific geometry of the system at hand. Nevertheless, when dealing with strongly confined fluids, it is not clear that hydrodynamics hold in the direction perpendicular to the confining walls. 

Kinetic theory and non-equilibrium statistical mechanics provide the appropriate context to investigate which is the right macroscopic description of transport under strong confinement,  providing also the expressions for the needed transport coefficient. Notice that, in principle, not only the values of the transport coefficient can be different from the bulk case, but also the structure of the transport equations itself can differ. A first issue to formulate a kinetic theory of strongly confined systems is to propose a well-founded kinetic equation for them. Up to now, to the best of our knowledge, most of the efforts in this direction have consisted in formulating model kinetic equations to investigated specific observed effects, mainly in the context of granular fluids \cite{BRyS13,SRyB14,BGMyB13,BBGyM16}.

In the present paper, a kinetic equation to study non-equilibrium processes in a confined quasi-one-dimensional dilute system of hard disks is formulated, and some of its fundamental properties are investigated. More specifically, the system is confined by two hard walls separated a distance between one and two particle diameters. The kinetic equation is derived by adapting the arguments leading to the Boltzmann equation for bulk systems \cite{DyvB77,RydL77} and, in particular, making use of the molecular chaos approximation. The main difference implied by the confinement is a restriction on the possible values of the scattering angle between colliding particles. This is the mathematical manifestation of the inhomogeneity introduced by the confinement. Of course, a first test to be passed by the kinetic equation is that it must reproduce the results obtained by means of the well stablished theory of inhomogeneous fluids under thermodynamic equilibrium conditions \cite{Ev92,SyL96,SyL97}. This includes not only the equilibrium density profile, but also the static pressure parameters associated to the perpendicular and parallel directions to the confining walls.

In the case of a strictly one-dimensional system of elastic hard rods, it is evident that the system does not approach a Maxwellian velocity distribution, given that particles just interchange their velocities in a collision. As a consequence, there is no H-theorem for such systems, contrary to what happens with a bulk system of hard disks (or spheres) \cite{DyvB77,RydL77}. A relevant question is, therefore, whether a Liapunov function and an associated $H$-theorem does exist for a quasi-one-dimensional system of hard disks, as the one described above and considered in this paper. If this is the case, does the H-function reduces to the equilibrium entropy, aside from proportionality and additive trivial constants?

In the next Section, the kinetic equation describing the dynamics of a dilute quasi-one-dimensional gas of hard disks is formulated. The arguments leading to the equation are essentially the same as those employed for a quasi-two-dimensional gas of hard spheres in ref. \cite{BGyM17}.  Also,  the boundary conditions to be employed when solving the equation are formulated. In Sec. \ref{s3}, an $H$-theorem implying the approach to equilibrium from any initial condition is proven.  The Liapunov $H$-function contains, in addition to the usual Boltzmann expression, a contribution associated with the confinement, which is a functional of the (non-equilibrium) density field. By means of the $H$ theorem, the unique stationary distribution is determined. As expected, it is a Gaussian in the velocity space with a non-uniform density field.  The latter is given as the solution of an integro-differential equation, whose exact solution seems hard to find in an analytical form. An expression valid in the limit of a very narrow channel (its width slightly larger that the diameter of the particles) is derived. 

From the quantity $H$, an expression for the entropy of the system is proposed in Sec. \ref{s4}.  When particularized for the equilibrium situation, the result agrees, in the appropriate limit, with the entropy computed by means of equilibrium statistical mechanics methods \cite{SyL96,SyL97}. Next, two pressure parameters, associated with the separation of the confined walls and with their length, respectively, are defined in terms of the entropy. The expression obtained for the former is consistent with the wall theorem for systems of hard spheres or disks \cite{Fi64,HyMc86,MGyB18}. 

As a simple application of the kinetic equation, in Sec. \ref{s5}, closed evolution equations for the two temperature parameters associated to the motion parallel and perpendicular to the confining walls, respectively, are derived,  for the simple case of homogeneous systems along the direction of the walls. The results are compared with molecular dynamics simulation data, and a good agreement is obtained. Finally Sec. \ref{s6} contains a summary of the conclusions and some final comments and perspectives.

\section{The kinetic equation}
\label{s2}
\begin{figure}
 \includegraphics[scale=0.6,angle=0]{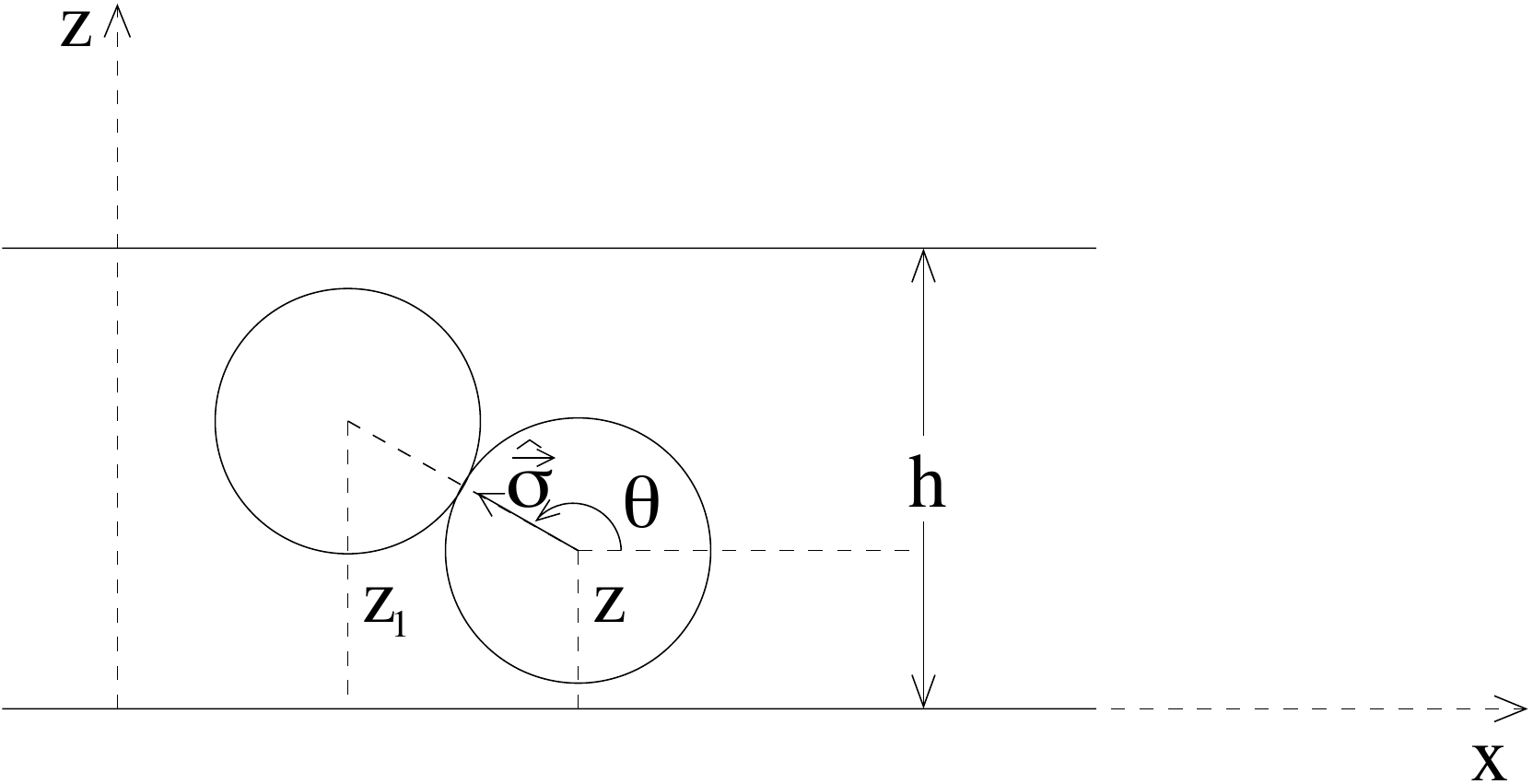}
\caption{(Color online) Collision between two hard disks in a quasi-one-dimensional confined system. The X and Z axis are taken as parallel and perpendicular to the confining plates, respectively. It is seen that the collision vector is univocally determined by the value of the polar angle $ \theta$.}
\label{fig1}
\end{figure}

A quasi-one-dimensional system of $N$ hard disks of mass $m$ and diameter $\sigma$ is considered. The particles are confined by two large, formally infinite, parallel straight hard walls separated a distance $h$, $\sigma < h< 2\sigma$. The number of particles per unit of area is assumed to be very low, so that the Stosszahlansatz or molecular chaos hypothesis \cite{DyvB77,RydL77}, assuming that there are no velocity correlations between particles before collisions, can be used. Moreover, it is also assumed that the one-particle distribution function of the gas of disks, $f({\bm r},{\bm v},t)$ can be taken as constant over displacements of the order of the diameter of the disks along the direction parallel to the hard walls. Trivially, the same property can not be assumed in the direction perpendicular to the walls, since the confinement breaks the isotropy of the dynamics and imposes relevant variations over distances smaller than the diameter of the particles when going from one hard wall to the other one. It follows that the relevant parameter defining the low density limit is the number of particles per unit of length along the direction parallel to the walls. A sketch of the system and the coordinates used to describe a collision is provided in Fig. \ref{fig1}. In mathematical language, the assumption mentioned above reads
\begin{equation}
\label{2.0}
f(x+\Delta x,z,{\bm v},t) \approx f(x,z,{\bm v},t),
\end{equation}
for $|\Delta x| = {\mathcal O}(\sigma)$. Then, following the usual arguments leading to the Boltzmann-Enskog equation, and reproduced in detail for a quasi-two-dimesional gas in \cite{BGyM17}, it is obtained
\begin{equation}
\label{2.1}
\left ( \frac{\partial}{\partial t} +{\bm v} \cdot  \frac{\partial}{\partial {\bm r}} \right) f({\bm r}, {\bm v},t)= J \left[ {\bm r},{\bm v}|f \right],
\end{equation}
with the collision term given by
\begin{equation}
\label{2.2}
J \left[ {\bm r},{\bm v}|f \right] = \sigma \int d{\bm v}_{1} \int_{\Omega_{\bm \sigma}(z)} d\widehat{\bm \sigma}\,   | {\bm g} \cdot \widehat {\bm \sigma}  | \left[  \Theta ({\bm g} \cdot \widehat {\bm \sigma}) b_{\bm \sigma}  - \Theta (-{\bm g} \cdot \widehat {\bm \sigma}) \right] f(x,z+\sigma_{z},{\bm v}_{1})f({\bm r},{\bm v},t) .
\end{equation}

Here, $\widehat{\bm \sigma}$ is the unit vector along the line joining the centers of the two particles at contact from particle $1$, $\sigma_{z} = \sigma \widehat{\sigma}_{z}$, ${\bm g} \equiv {\bm v}_{1} -{\bm v}$, ${\bm r} \equiv \left\{x,z \right\}$, $\Theta$ is the Heaviside step function, and $b_{\bm \sigma}$ is an operator changing all the velocities ${\bm v}$ and ${\bm v}_{1}$ to its right into the post-collisional values given by
\begin{equation}
\label{2.3}
{\bm v}^{\prime} \equiv  b_{\bm \sigma} {\bm v} = {\bm v}+ \left( {\bm g} \cdot \widehat{\bm \sigma} \right) \widehat{\bm \sigma},
\end{equation}
\begin{equation}
\label{2.4}
{\bm v}^{\prime}_{1} \equiv  b_{\bm \sigma} {\bm v}_{1} = {\bm v}_{1}- \left( {\bm g} \cdot \widehat{\bm \sigma} \right) \widehat{\bm \sigma}.
\end{equation}
The angular integration over $d\widehat{\bm \sigma}$ is restricted to the domain ${\Omega_{\bm \sigma}(z)}$ that is compatible with the confinement imposed by the parallel walls. In Fig. \ref{fig1} it is seen that the domain of integration depends on the coordinate $z$ of the target particle. Moreover, it is
\begin{equation}
\label{2.5}
\widehat{\bm \sigma} \equiv \left\{ \cos \theta, \sin \theta \right\}
\end{equation}
and $d\widehat {\bm \sigma}= d \theta$, where $\theta$ is the polar angle indicated in the figure. For fixed $z$, the coordinates $z_{1}$ and $\theta$ are related by
\begin{equation}
\label{2.6}
\sin \theta_{1} = \frac{z_{1}-z}{\sigma}, \quad \cos \theta_{1} = + \sqrt{1- \left( \frac{z_{1}-z}{\sigma} \right)^{1} } \quad \mbox{for}  \quad -\frac{\pi}{2} <\theta_{1} < \frac{\pi}{2},
\end{equation} 
\begin{equation}
\label{2.7}
\sin \theta_{2} = \frac{z_{1}-z}{\sigma}, \quad \cos \theta_{2} = - \sqrt{1- \left( \frac{z_{1}-z}{\sigma} \right)^{1} } \quad  \mbox{for}  \quad \frac{\pi}{2} <\theta_{2} < \frac{ 3 \pi}{2}.
\end{equation} 
It is convenient to decompose the collision term $J\left[{\bm r}, {\bm v}|f \right]$ given in Eq.\ (\ref{2.2}) into two parts corresponding each of them to one of the two intervals of values of $\theta$ identified above. Then, it is possible to make a change of variables from $\theta$ to $z_{1}$ in each of the two contributions, with the result
\begin{eqnarray}
\label{2.8}
J[{\bm r},{\bm v}|f] & = & \int d {\bm v}_{1} \int_{\sigma/2}^{h-\sigma/2} dz_{1}\, (\cos \theta_{1})^{-1} \left( 1+ {\mathcal P}_{{\bm \sigma}_{1}, {\bm \sigma}_{2}} \right) |{\bm g} \cdot \widehat{\bm \sigma}_{1}| \left[  \Theta ({\bm g} \cdot \widehat {\bm \sigma}_{1}) b_{\bm \sigma_{1}}  - \Theta (-{\bm g} \cdot \widehat {\bm \sigma}_{1}) \right] \nonumber \\
&& \times f(x,z_{1},{\bm v}_{1},t)f({\bm r},{\bm v},t),
\end{eqnarray}
where ${\mathcal P}_{{\bm \sigma}_{1}, {\bm \sigma}_{2}} $ is an operator permuting all the vectors $\widehat{\bm \sigma}_{1}$ (corresponding to an angle $\theta_{1}$ as defined in Eq. (\ref{2.6}))  and $\widehat{\bm \sigma}_{2}$  (corresponding to an angle $\theta_{2}$ as defined in Eq. (\ref{2.7})) to its right. The latter are defined by Eq. (\ref{2.5}) using the expressions given in Eqs. (\ref{2.6}) and (\ref{2.7}), respectively. 

The kinetic equation formulated in Eqs. (\ref{2.1}) and (\ref{2.8}) must be solved with the appropriate boundary conditions, namely \cite{BGyM17},
\begin{equation}
\label{2.9}
\Theta (v_{z}) f({\bm r},{\bm v},t) \delta \left( z-\frac{\sigma}{2} \right) = \Theta (v_{z}) f( {\bm r},{\bm v}^{*},t) \delta \left( z-\frac{\sigma}{2} \right),
\end{equation}
\begin{equation}
\label{2.10}
\Theta (- v_{z}) f({\bm r},{\bm v},t) \delta \left( z-h+ \frac{\sigma}{2} \right) = \Theta (- v_{z}) f( {\bm r},{\bm v}^{*},t) \delta \left( z-h+ \frac{\sigma}{2} \right),
\end{equation}
being
\begin{equation}
\label{2.11}
{\bm v}^{*} = {\bm v} -2 v_{z} \widehat{\bm e}_{z},
\end{equation}
where $\widehat{\bm e}_{z}$ is the unit vector along the positive direction of the $Z$ axis.

A useful property of the collision term (\ref{2.8}) is that for any arbitrary function of the velocity $\xi ({\bm v})$ it is
\begin{eqnarray}
\label{2.12}
\int d{\bm v}\, \xi ({\bm v}) J [{\bm r},{\bm v}|f] & = & \int d {\bm v} \int d {\bm v}_{1} \int_{\sigma/2}^{h-\sigma/2} dz_{1}\, \left( \cos \theta_{1} \right)^{-1} \left( 1+ {\mathcal P}_{{\bm \sigma}_{1}, {\bm \sigma}_{2}} \right) \left[ \xi( b_{{\bm \sigma}_{1}} {\bm v} )- \xi ({\bm v}) \right] \nonumber \\
&& \times |{\bm g} \cdot \widehat{\bm \sigma}_{1} | \Theta \left(- {\bm g} \cdot \widehat{\bm \sigma}_{1} \right) f(x,z_{1},{\bm v}_{1},t)f({\bm r},{\bm v},t).
\end{eqnarray}
As a consequence, for a function of the position and the velocity, $\xi({\bm r},{\bm v})$, one has
\begin{eqnarray}
\label{2.13}
\int d{\bm r} \int d{\bm v}\, \xi ({\bm r},{\bm v}) J [{\bm r},{\bm v}|f] & = & \frac{1}{2} \int dx  \int d {\bm v} \int d {\bm v}_{1} \int_{\sigma/2}^{h-\sigma/2} dz \int_{\sigma/2}^{h-\sigma/2} dz_{1}\, \left( \cos \theta_{1} \right)^{-1} \left( 1+ {\mathcal P}_{{\bm \sigma}_{1}, {\bm \sigma}_{2}} \right)  \nonumber \\
&& \times \left[ \xi( {\bm r},b_{{\bm \sigma}_{1}} {\bm v} ) + \xi( x,z_{1},b_{{\bm \sigma}_{1}} {\bm v}_{1} ) -  \xi ({\bm r}, {\bm v})  \xi (x,z_{1} ,{\bm v}_{1} ) \right] \nonumber \\
&& \times  |{\bm g} \cdot \widehat{\bm \sigma}_{1} | \Theta \left(- {\bm g} \cdot \widehat{\bm \sigma}_{1} \right) f(x,z_{1},{\bm v}_{1},t)f({\bm r},{\bm v},t).
\end{eqnarray}

In terms of the one-particle distribution function, the macroscopic fields of number density of particles, $n({\bm r},t)$, velocity flow, ${\bm u}({\bm r},t)$, and temperature, $T({\bm r},t)$, are defined in the usual way,
\begin{equation}
\label{2.14}
n({\bm r},t) \equiv \int d{\bm v}\ f({\bm r},{\bm v},t),
\end{equation}
\begin{equation}
\label{2.15}
n({\bm r},t) {\bm u}({\bm r},t) \equiv \int d{\bm v}\ {\bm v} f({\bm r},{\bm v},t),
\end{equation}
\begin{equation}
\label{2.16}
n({\bm r},t)k_{B} T({\bm r},t)  \equiv \int d{\bm v}\ \frac{1}{2} m \left[ {\bm v}-{\bm u} ({\bm r},t) \right]^{2}f({\bm r},{\bm v},t).
\end{equation}
In the last expression, $k_{B}$ is the Boltzmann constant. Trivially, the above definitions do not presuppose the existence of a two-dimensional hydrodynamic description. Integration of the kinetic equation  (\ref{2.1}) over the velocity  and use of Eq. (\ref{2.13}) gives the continuity equation
\begin{equation}
\label{2.17}
\frac{\partial n({\bm r},t)}{\partial t} = - \frac{\partial}{\partial {\bm r}}\cdot \left[ n({\bm r},t) {\bm u}({\bm r},t) \right],
\end{equation}
which is just the balance equation for the number of particles.
\section{The H theorem and the equilibrium distribution function}
\label{s3}
Let us define a global property of the system $H(t)$ by \cite{BMyG16}
\begin{equation}
\label{3.1}
H(t) \equiv H^{k}(t)+H^{c}(t),
\end{equation}
where the ``kinetic part'' $H^{k}(t)$ is given by
\begin{equation}
\label{3.2}
H^{k}(t) \equiv \int d {\bm r} \int d{\bm v} f({\bm r},{\bm v},t) \left[ \ln f({\bm r},{\bm v},t) -1 \right],
\end{equation}
while the ``configurational'' component is
\begin{equation}
\label{3.3}
H^{c}(t) \equiv \int d{\bm r} \int d {\bm r}_{1}\ n({\bm r},t) n({\bm r}_{1},t) \Theta \left( \sigma- | {\bm r}_{1} - {\bm r} | \right).
\end{equation}

The term $H^{k}(t)$ is the same as considered by Boltzmann in his celebrated $H$ theorem \cite{DyvB77,RydL77}. The configurational part is due to the confinement of the fluid between the two parallel plates. Its form is consistent with considering a local equilibrium approximation for the $N$ particle distribution function \cite{Re78a,Re78b,MGyB18}, and also with the expression for the equilibrium entropy of a system of hard disks in the second virial coefficient approximation \cite{KyD71, HyMc86}. Our aim is to study the time evolution of $H(t)$ in the confined system. Consider first the time derivative of the kinetic part,
\begin{equation}
\label{3.4}
\frac{\partial H^{k}(t)}{\partial t} = \int d{\bm r} \int d{\bm v}  \ln f({\bm r},{\bm v},t) \frac{\partial f({\bm r},{\bm v},t)}{\partial t}\, .
\end{equation}
The term coming from the free flow contribution in the kinetic equation has the form
\begin{eqnarray}
\label{3.5}
-\int d{\bm r} \int d{\bm v}\, \left[  \ln f({\bm r},{\bm v},t) \right] {\bm v} \cdot \frac{\partial f}{\partial {\bm r}} & = & - \int d{\bm r} \int d{\bm v}\, \frac{\partial}{\partial {\bm r}}\, \cdot \left[ {\bm v} \left( f \ln f-f \right) \right] \nonumber \\
&=& - \int d{\bm v} \int_{S} d{\bm S} \cdot {\bm v}  \left( f \ln f-f \right)=0,
\end{eqnarray}
where the Gauss theorem has been used and the property  the flux of any property through the boundaries $S$ vanishes has been employed. Then, Eq. (\ref{3.4}) reduces to
\begin{equation}
\label{3.6}
\frac{\partial H^{k}(t)}{\partial t} = \int d{\bm r} \int d{\bm v}   \ln f({\bm r},{\bm v},t)  J [ {\bm r},{\bm v}|f].
\end{equation}
By means of the general property (\ref{2.13}), one gets
\begin{eqnarray}
\label{3.7}
\frac{\partial H^{k}(t)}{\partial t}& =& \frac{1}{2} \int dx  \int d {\bm v} \int d {\bm v}_{1} \int_{\sigma/2}^{h-\sigma/2} dz \int_{\sigma/2}^{h-\sigma/2} dz_{1}\, \left( \cos \theta_{1} \right)^{-1} \left( 1+ {\mathcal P}_{{\bm \sigma}_{1}, {\bm \sigma}_{2}} \right)  \nonumber \\
&& \times \ln \frac{f({\bm r}, b_{{\bm \sigma}_{1}}{\bm v} ,t) f(x,z_{1},b_{{\bm \sigma}_{1}}{\bm v}_{1} ,t)}{f({\bm r},{\bm v},t) f(x,z_{1},{\bm v}_{1},t)}\,  |{\bm g} \cdot \widehat{\bm \sigma}_{1} | \Theta \left(- {\bm g} \cdot \widehat{\bm \sigma}_{1} \right) f(x,z_{1},{\bm v}_{1},t)f({\bm r},{\bm v},t). \nonumber \\
\end{eqnarray}
Next, the inequality $x ( \ln y -\ln x ) \leq y-x$, valid for $x,y>0$ is employed. The equality sign only holds for $x=y$. It is obtained 
\begin{eqnarray}
\label{3.8}
\frac{\partial H^{k}(t)}{\partial t}&\leq &  \frac{1}{2} \int dx  \int d {\bm v} \int d {\bm v}_{1} \int_{\sigma/2}^{h-\sigma/2} dz \int_{\sigma/2}^{h-\sigma/2} dz_{1}\, \left( \cos \theta_{1} \right)^{-1} \left( 1+ {\mathcal P}_{{\bm \sigma}_{1}, {\bm \sigma}_{2}} \right)  
\nonumber \\
&& \times \left[ f({\bm r}, b_{{\bm \sigma}_{1}}{\bm v} ,t) f(x,z_{1},b_{{\bm \sigma}_{1}} {\bm v}_{1},t) - f({\bm r},{\bm v},t) f(x,z_{1},{\bm v}_{1},t) \right]  |{\bm g} \cdot \widehat{\bm \sigma}_{1} | \Theta \left(- {\bm g} \cdot \widehat{\bm \sigma}_{1} \right) \nonumber \\
&=& 2 \int dx   \int_{\sigma/2}^{h-\sigma/2} dz \int_{\sigma/2}^{h-\sigma/2} dz_{1}\, (\tan \theta_{1})   n({\bm r},t) n(x,z_{1},t) u_{z} (x,z_{1},t).
\end{eqnarray}
In the last transformation we have made the change $b_{{\bm \sigma}_{1}} {\bm v} , b_{{\bm \sigma}_{1}} {\bm v}_{1} \rightarrow {\bm v}, {\bm v}_{1}$, to realize that the four terms on the right hand side are the same. Afterwards, the definition of the macroscopic velocity field, Eq. (\ref{2.15}) has been used. 

To study the time evolution of the configurational component of $H$ defined in Eq.\ (\ref{3.3}), the coordinates $\zeta \equiv   
|{\bm r}–{\bm r}_{1}| $ and $z_{1}$ will be employed for ${\bm r}_{1}$. Then, it is found
\begin{equation}
\label{3.9}
H^{c}(t) = \int dx   \int_{\sigma/2}^{h-\sigma/2} dz \int_{\sigma/2}^{h-\sigma/2} dz_{1} \int_{|z-z_{1}|}^{\infty} d\zeta \, \left( \cos \varphi \right)^{-1} n({\bm r},t) n(x,z_{1},t) \Theta (\sigma- \zeta),
\end{equation}
where
\begin{equation}
\label{3.10}
\cos \varphi \equiv + \sqrt{ 1- \left(  \frac{z_{1}-z}{\zeta} \right)^{2}}.
\end{equation}
Carrying out the integration over $\zeta$ yields
\begin{equation}
\label{3.11}
H^{c}(t) = \int dx   \int_{\sigma/2}^{h-\sigma/2} dz \int_{\sigma/2}^{h-\sigma/2} dz_{1}\,   n({\bm r},t) n(x,z_{1},t) \sqrt{ \sigma^{2}- (z-z_{1} )^{2} }
\end{equation}
and, from here,
\begin{eqnarray}
\label{3.12}
\frac{\partial H^{c}(t)}{\partial t} &=& \int dx   \int_{\sigma/2}^{h-\sigma/2} dz \int_{\sigma/2}^{h-\sigma/2} dz_{1}\,  \left[ \frac{\partial n({\bm r},t)}{\partial t}\,  n(x,z_{1},t) + n({\bm r},t) \frac{\partial n(x,z_{1},t)}{\partial t}\, \right]  \nonumber \\
&& \times \sqrt{ \sigma^{2}- (z-z_{1} )^{2} }.
\end{eqnarray}
Next, the continuity equation (\ref{2.17}) is used to rewrite the above equation as 
\begin{equation}
\label{3.13}
\frac{\partial H^{c}(t)}{\partial t} = \int dx   \int_{\sigma/2}^{h-\sigma/2} dz \int_{\sigma/2}^{h-\sigma/2} dz_{1}\, n({\bm r},t) n(x,z_{1},t) u_{z} ( {\bm r},t) \frac{\partial}{\partial z} \sqrt{ \sigma^{2}- (z-z_{1} )^{2} }.
\end{equation}
To derive this expression, an integration by parts has been performed. Moreover, the assumed condition formulated by Eq. (\ref{2.1}) has been employed. More precisely, it has been used for $\Delta x = \sqrt{\sigma^{2} -(z-z_{1})^{2}} \leq \sigma$. It is just a matter os simple transformations to convert Eq. (\ref{3.13}) into
\begin{equation}
\label{3.14}
\frac{\partial H^{c}(t)}{\partial t} = -2 \int dx   \int_{\sigma/2}^{h-\sigma/2} dz \int_{\sigma/2}^{h-\sigma/2} dz_{1}\, (\tan \theta_{1}) n({\bm r},t) n(x,z_{1},t) u_{z} ( x,z_{1},t) .
\end{equation}
Use of Eqs.\ (\ref{3.1}), (\ref{3.8}), and (\ref{3.14}) yields
\begin{equation}
\label{3.15}
\frac{\partial H(t)}{\partial t} = \frac{\partial H^{k}(t)}{\partial t} +\frac{\partial H^{(c)}(t)}{\partial t} \leq 0.
\end{equation}
The equality sign only applies if (see below Eq. (\ref{3.7}))
\begin{equation}
\label{3.16}
 f({\bm r}, b_{{\bm \sigma}_{i}}{\bm v} ,t) f(x,z_{1},b_{{\bm \sigma}_{i}} {\bm v}_{1},t) = f({\bm r},{\bm v},t) f(x,z_{1},{\bm v}_{1},t),
 \end{equation}
 for both $i=1$ and $i=2$, and all the allowed values of $x, {\bm v}, {\bm v}_{1}, z$, and  $z_{1}$. 
 
 At this point, it is interesting to consider what happens in the limit of a true one-dimensional system, i.e., when $h \rightarrow \sigma$. Then, for initial conditions in which all the particles are inside the system, the distribution function at later times has the form
 \begin{equation}
 \label{3.17}
 f({\bm r},{\bm v},t)= \delta \left(z-\frac{\sigma}{2} \right) \delta (v_{z}) \widetilde{f} (x,v_{x},t),
 \end{equation}
and, in all collisions, the angle $\theta$ given in Eqs. (\ref{2.6}) and (\ref{2.7}) is either $\theta_{1}=0$ or $\theta_{2}= \pi$, implying that $\widehat{\sigma}_{x}=1$ or $\widehat{\sigma}_{x}=-1$, respectively. Then, the well known result that in elastic one-dimensional collisions, particles just interchange their velocities is recovered. As a consequence, Eq.\ (\ref{3.16}) is just a trivial identity, implying no condition on the form of the one-particle distribution function, and $H(t)=H(0)$ for any initial condition. Notice that the collision term also identically vanishes in this limit and the time evolution of the one-particle distribution function is fully determined by the free flow part of the kinetic equation.

Outside the above limiting case, following quite standard reasonings it is easily shown that the only physically relevant stationary solution of the kinetic equation is  \cite{UyF63,BGyM17}
\begin{equation}
\label{3.18}
f_{st}({\bm r},{\bm v})= n(z) \varphi_{MB}({\bm v}),
\end{equation}
where $\varphi_{MB}({\bm v})$ is the two-dimensional Maxwell-Boltzmann velocity distribution,
\begin{equation}
\label{3.19}
\varphi_{MB}({\bm v}) = \frac{m}{2 \pi k_{B} T} \exp \left[- \frac{mv^{2}}{2 k_{B} T} \right] \, .
\end{equation}
The density profile along the $z$ direction $n(z)$ is identified by substituting Eq. (\ref{3.18}) into the kinetic equation Eq. (\ref{2.1}) and requiring stationarity. In this way, it is obtained that the density field must verify
\begin{equation}
\label{3.20}
\frac{\partial \ln n(z)}{\partial z} = -2 \int_{ \sigma / 2}^{h-\sigma/2} dz_{1} \tan \theta_{1} n(z_{1})\, .
\end{equation}
Here $\theta_{1}$ is the angle defined in Eq. (\ref{2.6}), i.e.,
\begin{equation}
\label{3.21}
\tan \theta_{1} \equiv \frac{z_{1}-z}{ \sqrt{\sigma^{2}- (z_{1}-z)^{2}}}\, ,
\end{equation}
 The above equation does not seen easy to solve analytically, but a useful approximation ca be found by considering the limit \cite{MGyB19},
 \begin{equation}
 \label{3.21a}
 \epsilon \equiv \frac{h-\sigma}{\sigma} \ll 1. 
 \end{equation}
 In this limit, for all the allowed values of $z$ and $z_{1}$ inside the system, it is $|z-z_{1}| /\sigma \ll 1$, and in the lowest order, $\tan \theta \approx (z_{1}-z)/ \sigma$. Then, Eq. (\ref{3.20}) simplifies to
 \begin{equation}
 \label{3.22}
 \frac{\partial \ln n(z)}{\partial z} = \frac{2}{\sigma} \int _{\sigma /2}^{h-\sigma/2} dz_{1} (z-z_{1})  n(z_{1})\, .
 \end{equation}
 Introduce the variables
 \begin{equation}
 \label{3.23}
 \eta  \equiv z-\frac{h}{2}\, ,  \quad  \eta_{1}  \equiv z_{1}-\frac{h}{2}\, .
   \end{equation}
 Then, by symmetry it must be $n(\eta)= n(-\eta)$ and the integral in Eq. (\ref{3.22}) can be easily performed to get
 \begin{equation}
 \label{3.24}
 \frac{\partial \ln n(\eta)}{\partial \eta} = \frac{2 n_{0}(h-\sigma)}{\sigma}\,  \eta ,
 \end{equation}
 where $n_{0}$ is the average number of particles per unit of area. The solution of this equation is
 \begin{equation}
 \label{3.25}
 n(z)= \frac{n_{0} (h-\sigma)}{b} \exp \left[a \left(z-\frac{h}{2} \right)^{2} \right]\, .
 \end{equation}
Here,
\begin{equation}
\label{3.26}
a \equiv n_{0} \left(  \frac{h}{\sigma} -1 \right)\, ,
\end{equation}
 \begin{equation}
 \label{3.27}
 b \equiv \sqrt{\frac{\pi}{a}} \erfi \left[ \frac{\sqrt{a} \left( h - \sigma \right)}{2} \right],
 \end{equation}
and the imaginary error function, $\erfi(x)$, is defined as
\begin{equation}
\label{3.28}
\erfi(x) \equiv \frac{1}{\sqrt{\pi}} \int_{-x}^{x} dx^{\prime} e^{x^{\prime 2}}\, .
\end{equation}
 The transversal  density profile given by eq.\ (\ref{3.25}) has the same qualitative form as the one reported in refs. \cite{BGyM17} and \cite{BMyG16} for the case of a quasi-two-dimensional gas of hard spheres confined between two parallel plates. Nevertheless,  while the latter is an exact consequence of the kinetic equation, and valid for any separation of the plates between $\sigma$ and $2 \sigma$, the profile (\ref{3.25}) has been obtained here in the limit $h \rightarrow \sigma^{+}$.
 
 \section{The entropy and the pressure parameters}
 \label{s4}
 The results reported in the previous section suggest defining an entropy $S$ for the system we are considering by
 \begin{equation}
 \label{4.1}
 S(t) \equiv -k_{B} \left[ H(t) + N \ln h_{0}^{2} \right],
 \end{equation}
 with $H(t)$ given by Eq.\ (\ref{3.1}) and $c_{0}$ being  an indeterminate constant with dimensions of action divided by mass, introduced to render dimensionless the argument of the logarithm. For an isolated system, the $H$-theorem implies that
 \begin{equation}
 \label{4.2}
 \frac{dS(t)}{dt} \geq 0,
 \end{equation}
 reaching the maximum value, $S_{st}$,  at the equilibrium stationary state. Using the decomposition in Eq. (\ref{3.1}), it is
 \begin{equation}
 \label{4.3}
 S(t)= S_{st}^{k}+ S_{st}^{c},
 \end{equation}
 where
 \begin{equation}
 \label{4.4}
 S_{st}^{k}\equiv  -k_{B} \int d {\bm r} \int d{\bm v} f_{st}({\bm r},{\bm v}) \left\{ \ln \left[ f_{st}({\bm r},{\bm v}) c_{0}^{2} \right] -1 \right\}
 \end{equation}
 and 
 \begin{equation}
 \label{4.5}
 S_{st}^{c} \equiv -k_{B} \int dx   \int_{\sigma/2}^{h-\sigma/2} dz \int_{\sigma/2}^{h-\sigma/2} dz_{1}\,   n(z) n(z_{1})\sqrt{ \sigma^{2}- (z-z_{1} )^{2} } .
 \end{equation}
 Here, Eq. (\ref{3.11}) has been employed. The exact evaluation of these expressions in an analytical form would require to know the exact density profile $n(z)$. Instead, the limit $\epsilon \equiv (h-\sigma)/\sigma \ll 1$ will be considered again. In this limit, Eq.\, (\ref{4.5}) is approximated by
 \begin{equation}
 \label{4.6}
 S_{st}^{c}= -k_{B} \sigma \int_{-h/2+\sigma/2}^{h/2-\sigma/2} d\eta \int_{-h/2+\sigma/2}^{h/2-\sigma/2} d\eta_{1} \, n(\eta) n(\eta_{1}) \left[ 1 - \frac{ (\eta-\eta_{1})^{2}}{2 \sigma^{2}} \right].
 \end{equation}
 The variables $\eta$ and $\eta_{1}$ have been defined in Eqs.\, (\ref{3.23}). Using the expression of the transversal density field given in Eq.\ (\ref{3.25}) it is straightforward to obtain
 \begin{equation}
 \label{4.7}
 S_{st}^{c} = -\frac{Nk_{B}}{2} \left[ 1+2 \sigma n_{0}(h-\sigma) \right].
 \end{equation}
 In the same approximation of $\epsilon \ll1$, the kinetic part of the equilibrium entropy is found to be
 \begin{equation}
 \label{4.8}
 S_{st}^{k} =  -Nk_{B} \left[ \ln \frac{n_{0}(h-\sigma)}{b} - \ln T + \ln \frac{m c_{0}^{2}}{2 \pi k_{B}}- \frac{7}{2} \right] \, .
 \end{equation}
 The quantity $b$ appearing in the last expression is given by Eq. (\ref{3.27}). The equilibrium entropy can also be decomposed, in a more transparent form, into an ideal part, $S_{st}^{id}$, and an excess part, $S_{st}^{ex}$,
 \begin{equation}
 \label{4.9}
 S_{st}= S_{st}^{id}+S_{st}^{ex},
 \end{equation}
 \begin{equation}
 \label{4.10}
 S_{st}^{id} = -Nk_{B} \left[ \ln  n_{0} - \ln T + \ln \frac{m}{2 \pi k_{B}}- \frac{5}{2} \right] \, ,
 \end{equation}
 \begin{equation}
 \label{4.11}
 S_{st}^{ex} = -N k_{B} \left( \ln \frac{h-\sigma}{b} +\frac{\sigma N}{L} \right).
 \end{equation}
 From the Helmholtz free energy $F \equiv E-TS$, where $E$ is the internal energy, two different pressure parameters of the system are defined \cite{BMyG16,SyL96,SyL97}. They are referred to as the lateral pressure,
 \begin{equation}
 \label{4.12}
 p_{lat} \equiv -h^{-1} \left( \frac{\partial F}{\partial L} \right)_{h,N,T} = \frac{T}{h} \left( \frac{\partial S}{\partial L} \right)_{h,N,T}\, ,
 \end{equation}
 and a transversal pressure,
 \begin{equation}
 \label{4.13}
 p_{trans} \equiv \left( \frac{\partial F}{\partial h} \right)_{L,N,T} = \frac{T}{L} \left( \frac{\partial S}{\partial h} \right)_{L,N,T}\, .
 \end{equation}
 The second equalities in the above two equations follow because the internal energy per particle of a system of hard disks only depends on the temperature. Using the expressions of the entropy given in Eqs.\ (\ref{4.9})-(\ref{4.11}), it is found
 \begin{equation}
 \label{4.14}
 p_{lat}= \frac{Nk_{B}T}{h} \left[ \frac{3}{2L}+ \frac{N \sigma}{L^{2}} -\frac{h-\sigma}{2N} n(h-\sigma/2) \right],
 \end{equation}
 \begin{equation}
 \label{4.15}
 p_{trans} = n(h-\sigma/2) k_{B} T.
 \end{equation}
 The last result, relating the transversal pressure to the density of the fluid at the walls by means of the ideal gas equation of state, is nothing else but the wall theorem for the pressure tensor \cite{Fi64} applied to the present system. Actually, it has been proven recently that the theorem applies not only in equilibrium situations, but also holds in arbitrary non-equilibrium states \cite{MGyB18}. Moreover, a similar wall theorem exists for the heat flux. On the other hand, the expression for the lateral pressure only reduces to the ideal gas equation of state in the bulk limit $N \rightarrow \infty$, $L \rightarrow \infty$, $h \rightarrow \infty$, and $n \rightarrow n_{0}= N/Lh = constant$.

 \section{Relaxation to equilibrium of a system with anisotropy in velocity space}
 \label{s5}
 
 As an application of the kinetic equation studied in the previous sections, the relaxation towards equilibrium of a system with an anisotropic inicial velocity distribution will be investigated in this Section. For simplicity, states homogeneous along the $x$ direction and with vanishing velocity flow will be considered. The evolution of the state of the system will be characterized by two  temperature parameters, $\overline{T}_{x}$ and $\overline{T}_{z}$, defined as averages  in the $z$ direction of the kinetic energy associated with $v_{x}$ ans $v_{y}$, respectively,
 \begin{equation}
 \label{5.1}
 n_{0}k_{B}\overline{T}_{x}= \frac{m}{h-\sigma}\, \int_{\sigma/2}^{h-\sigma/2} dz \int d{\bm v}\, v_{x}^{2} f(z,{\bm v},t),
 \end{equation}
 \begin{equation}
 \label{5.2}
 n_{0}k_{B} \overline{T}_{z}= \frac{m}{h-\sigma}\, \int_{\sigma/2}^{h-\sigma/2} dz \int d{\bm v}\, v_{z}^{2} f(z,{\bm v},t).
 \end{equation}
 An average temperature, $\overline{T}$, can be defined in terms of the two above temperature parameters as 
 \begin{equation}
 \label{5.3}
 \overline{T} \equiv \frac{ \overline{T}_{x}+ \overline{T}_{z}}{2}\, .
\end{equation}
 From the kinetic equation (\ref{2.1}), using the elastic boundary conditions (\ref{2.9}) and (\ref{2.10}), evolution equations for the temperature parameters are found,
 \begin{equation}
 \label{5.4}
 n_{0}k_{B} \frac{\partial \overline{T}_{x}}{\partial t} = \frac{m}{h-\sigma} \int_{\sigma/2}^{h-\sigma/2} dz \int d{\bm v} v_{x}^{2} J[z,{\bm v}|f],
 \end{equation}
 \begin{equation}
 \label{5.5}
  n_{0}k_{B} \frac{\partial \overline{T}_{z}}{\partial t} = \frac{m}{h-\sigma} \int_{\sigma/2}^{h-\sigma/2} dz \int d{\bm v} v_{y}^{2} J[z,{\bm v}|f].
  \end{equation}
 By using the property given in Eq.\ (\ref{2.13}), it is easily seen that
 \begin{equation}
 \label{5.6}
 \frac{\partial \overline{T}}{\partial t} =0,
 \end{equation}
 as required by kinetic energy conservation. A closed system of equations is obtained from Eqs.\ (\ref{5.4}) and (\ref{5.5}) if the one particle distribution function only depends on time through $\overline{T}_{x}$ and $\overline{T}_{z}$. Here, a local-equilibrium-like approximation will be employed, namely
 \begin{equation}
  \label{5.7}
  f(z,{\bm v},t) \approx \frac{n_{0}}{\pi \omega_{x}(t) \omega_{z}(t)} \exp \left[  - \left( \frac{v_{x}}{\omega_{x}(t)} \right)^{2}- 
 \left( \frac{v_{x}}{\omega_{x}(t)} \right)^{2} \right],
 \end{equation}
 where
 \begin{equation}
 \label{5.8}
 \omega_{x}^{2}(t) \equiv \frac{2k_{B}\overline{T}_{x}(t)}{m},
 \end{equation}
 \begin{equation}
 \label{5.9}
 \omega_{z}^{2}(t) \equiv \frac{2k_{B}\overline{T}_{z}(t)}{m}.
 \end{equation}
 Notice that the approximation given by Eq.\, (\ref{5.7}), in addition to a Gaussian distribution in velocity space, also assumes 
homogeneity in the $z$ direction. This is expected to be an accurate approximation at low densities, since in this case we saw that the dependence of the density on $z$ is very weak at equilibrium. Substitution of Eq. (\ref{5.7} into Eqs. (\ref{5.4}) and (\ref{5.5}), and use of Eq.\ (\ref{2.13}), after some algebra yields to equations involving rather complicated integrations. Here results using the symbolic computation program Matlab  \cite{Matlab} will be reported. They correspond to an expansion in the 
parameter $\epsilon$ defined in Eq.\ (\ref{3.21a}) up to the first nontrivial order, that is $\epsilon^{3}$,
\begin{equation}
\label{5.10}
\frac{d\overline{T}_{z}}{ds} = \frac{4 \epsilon^{3}}{3} \left[ \overline{T}- \overline{T}_{z}(s) \right],
\end{equation}
where the dimensionless time scale $s$ defined by
\begin{equation}
\label{5.11}
s(t) \equiv \frac{n \sigma \sqrt{2}}{\sqrt{\pi}} \int_{0}^{t} dt\, \omega_{x}(t),
\end{equation}
has been introduced. It is proportional to the average number of collisions per particle in the time interval between $0$ and $t$.
The equation for $\overline{T}_{x}(s)$ follows by using $\overline{T}_{x} = 2 \overline{T}-\overline{T}_{z}$, taking into account that $\overline{T}$ is constant. The general solution of Eq. (\ref{5.10}) is
\begin{equation}
\label{5.12}
\overline{T}_{z}(t) = \overline{T} \left[ 1-C \exp \left( -\frac{4 \epsilon^{3} s}{3} \right) \right],
\end{equation}
$C$ being some constant to be determined from the initial condition. Of course, because of the kinetic energy conservation, it is
\begin{equation}
\label{5.12}
\overline{T}_{x}(t) = \overline{T} \left[ 1+C \exp \left( -\frac{4 \epsilon^{3} s}{3} \right) \right].
\end{equation}
It is seen that in the long time limit, it is $\overline{T}_{x}=\overline{T}_{z} = \overline{T}$, in agreement with the $H$ theorem derived in Sec. \ref{s3}. The relaxation of the temperature parameter is exponential with a characteristic relaxation time $\tau$ that, in the dimensionless time scale $s$ is given by
\begin{equation}
\label{5.13}
\tau = \frac{3}{4 \epsilon^{3}}.
\end{equation}
It follows that the relaxation is slower the narrower  the system. This can be easily understood, since as  the separation of the two walls decreases, the collisions between particles become more frontal (the range of allowed values of the angle $\theta$ in Fig. \ref{fig1} becomes narrower around $\theta =0$ and $\theta = \pi$), and the energy change  in collisions also gets smaller, leading to a less efficient mechanism of approaching equilibrium.

\begin{figure}
\includegraphics[scale=0.4
,angle=0]{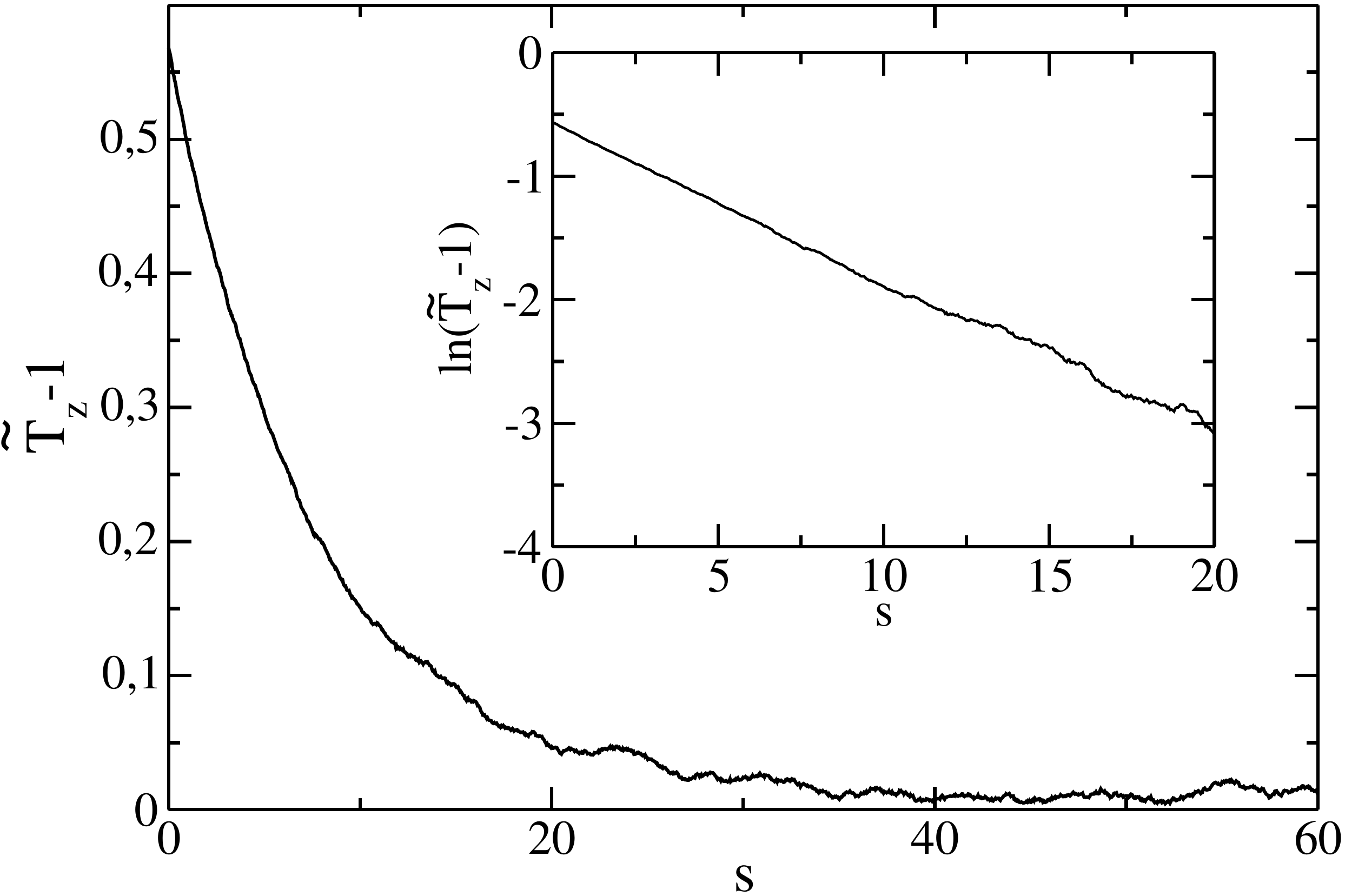}
\caption{Molecular Dynamics simulation results for the decay of the scaled averaged vertical temperature parameter, $\widetilde{T}_{z} \equiv (\overline{T}_{z}-T)/T)$, where $T$ is the homogeneous temperature, that remains constant in time. Time is measured in the dimensionless units defined in Eq.\ (ref{5.11}).  The separation of the two hard walls is $h=1,5 \sigma$, the initial temperature parameters  are $T_{z}(0)= 5 T_{x}(0)$, and the number density $N=0,06 \sigma^{-2}$. An exponential behavior in clearly identified in  the logarithmic representation used in the inset.}
\label{fig2}
\end{figure}

\begin{figure}
\includegraphics[scale=0.4,angle=0]{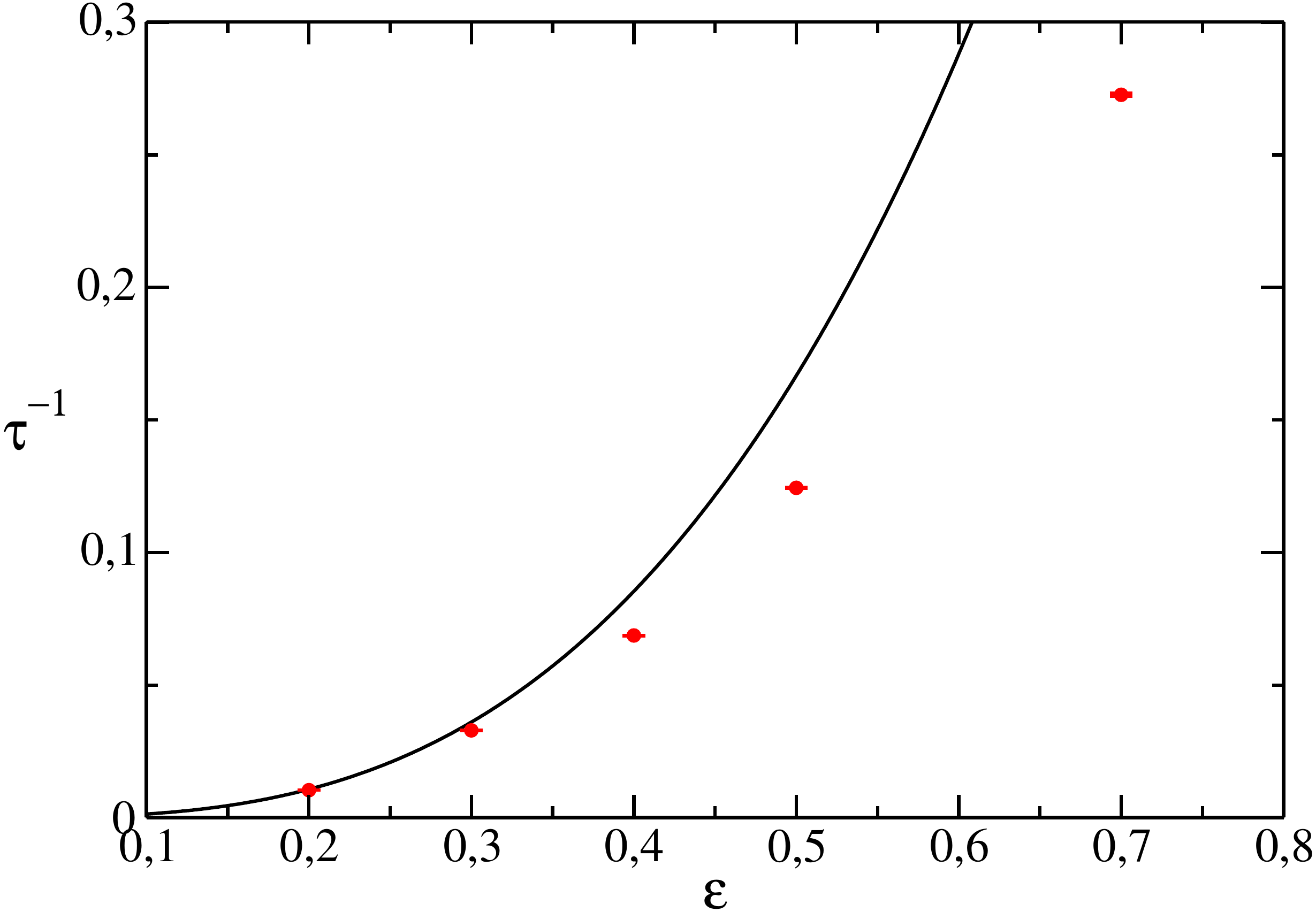}
\caption{Inverse of the characteristic relaxation time,  $\tau$, of the temperature parameters in a confined quasi-one-dimensional systems of hard discs, as a function of the dimensionless parameter $\epsilon$, defined in Eq. (\ref{3.21a}), that is a measure of the width of the system. Al the other parameters have the same values as in Fig. \ref{fig2}. Time is measured in units of $s$ as defined in Eq. (\ref{5.11}). The (red) symbols have been obtained by fitting the MD relaxation curves to exponentials, as described in the main text. The solid (black) lines are the theoretical prediction derived here, Eq. (\ref{5.13}). The system was initially prepared with anisotropy in velocity space, as described in the main text.}
\label{fig3}
\end{figure}

To investigate the accuracy of the above theoretical predictions, Molecular Dynamics simulations have been performed. The results to be reported in the following correspond to a system of $500$ hard disks, and the number density is $n_{0}=0,06 \sigma^{-2}$. Moreover, periodic boundary conditions have been employed in the $x$ direction. The initial velocity distribution was chosen homogeneous  and Gaussian for both components, with initial temperature parameters $T_{z}(0)= 5 T_{x}(0)$. As an example, in Fig.\ \ref{fig2} the time evolution of the normalized vertical temperature parameter $\widetilde{T}_{z} \equiv (\overline{T}_{z}-T)/T$,  is plotted as a function of time, measured in the  scale $s$ defined in Eq.\ (\ref{5.11}). In the inset, a logarithm plotted is shown, so that the exponential shape of the decay be identified. Similar results were found when varying the value of $\epsilon \equiv (h-\sigma)/\sigma$, keeping the same the remaining parameters of the system and the initial conditions. In Fig.\ \ref{fig3}, the simulation results for the inverse of the relaxation time $\tau^{-1}$ obtaining by fitting the decays to  $\widetilde{T}_{z}(s) = \widetilde{T}_{z}(0) \exp (-s/\tau)$, are compared with the theoretical prediction given in Eq. (\ref{5.13}), i.e. $\tau^{-1}= 4 \epsilon^{3} /3$. The agreement is very good for small values of $\epsilon$, but it clearly gets worse as the distance between the two confining walls increases. This was to be expected, since the theoretical prediction has being derived in the limit $\epsilon \ll 1$.

\section{Conclusions and final comments}
\label{s6}
In this paper, a dilute system of hard disks strongly confined between two hard walls has been considered. A Boltzmann-like kinetic equation has been derived. The effect of the strong confinement is to impose a limitation on  the possible collisional angles. It has been shown that the kinetic equation admits an entropy Lyapunov function describing a monotonic approach to equilibrium. This result strongly suggests the existence of hydrodynamics for confining quasi-one-dimensional systems. This is a nontrivial expectation, since strictly one-dimensional systems do not admit an hydrodynamic description, following that the one-dimensional system limit would be singular.  Consistently, it has been discussed why the analysis leading to the equilibrium approach fails in the limiting case of a one-dimensional system.

The equilibrium distribution function is Gaussian in velocity space, and exhibits nonuniform density in the direction perpendicular to the confining walls. An analytical expression has been obtained for this density profile in the limit of very strong confinement, i.e. when the separation between the walls in slightly larger than the diameter of the particles. The form of the equilibrium distribution is consistent with the result following from equilibrium statistical mechanics. Next, the expression of the equilibrium entropy has been derived and, from it, the values of two pressure parameters associated to the directions perpendicular and parallel to the confining walls, respectively. These two parameters differ, as a consequence of the anisotropy induced by the confinement. 

As a simple application of the kinetic equation, the time evolution of the temperature parameters associated to the vertical and horizontal degrees of freedom have been investigated in a system that is homogeneous in the horizontal direction. To get closed equations for the two parameters, it has been assumed that the velocity distribution function is Gaussian at all times with no correlations, although with different second moments of the longitudinal and transversal components. In the limit of very strong confinement mentioned above, the equations for the temperature parameters can be solved analytically, and an exponential relaxation towards the (equal) equilibrium values is obtained, when a time scale associated to the cumulative number of collisions per particle is employed. This allows the direct identification of a characteristic relaxation time. Molecular dynamics simulations have been performed, and a good agreement with the theoretical predictions has been found.

There are several future perspectives of the research initiated in this paper. The first and the most natural, seems to be to derive macroscopic transport equation for the density, velocity field, and temperature, i.e., hydrodynamic equations. It is not easy to guess a priori which is the effect of the confinement as compared with the case of bulk systems. Not only the expression of the transport coefficients will be different, by sure, but it is possible that the structure of the equation itself be modified, for instance with the presence of new terms. 

Having in mind transport in nanofluids, it is clear that another interesting extension of the present analysis  is to consider inelastic hard disks, as a simple model of granular gases. The kinetic Boltzmann equation considered here can be easily adapted for inelastic collisions, similarly to what has been done some time ago for the ordinary, no confined, Boltzmann equation  \cite{GyS95}. Notice that to maintain a granular gas in some macroscopic fluidized state, it is necessary to continuously inject energy into the system, something that can be done, for instance, by vibrating one or the two walls confining the system. This breaks the isotropy in velocity space and renders relevant to distinguish  between the two temperature parameters \cite{MGyB19}, as carried out  in the previous section. The next task will be to derive transport equations that take into account both, the effect of the strong confinement and the effect of the inelasticity in collisions. It is already known how to deal with the  latter, either by using linear response  theory \cite{DyB02,ByR04} or a modified Chapman-Enskog expansion \cite{BDKyS98,ByC01}. The knowledge about the consequences of the confinement is just the future 
perspective pointed out in the previous paragraph.

\begin{acknowledgements}
This research was supported by the Ministerio de Econom\'ia, Industria y Competitividad (Spain) through Grant No. FIS2017-87117-P (partially financed by FEDER funds).
\end{acknowledgements}

\end{document}